\documentclass[aps,preprint,superscriptaddress,amsmath,amssymb]{revtex4}
\usepackage{graphicx,color,slashed}
\usepackage{subfigure,bbold}

\def\be{\begin{eqnarray}}
\def\ed{\end{eqnarray}}
\def\non{\nonumber}

\begin{document}

{\begin{flushright}{KIAS-P15066}
\end{flushright}}

\title{\fontsize{13.2pt}{0pt}\selectfont Higgs singlet boson as a diphoton resonance in a vector-like quark model  }

\author{  R.~Benbrik \footnote{Email: rbenbrik@ictp.it}}
\affiliation{LPHEA, Semlalia, Cadi Ayyad University, Marrakech, Morocco}
\affiliation{MSISM Team, Facult\'e Polydisciplinaire de Safi, Sidi Bouzid B.P 4162, 46000 Safi, Morocco}

\author{ Chuan-Hung Chen \footnote{Email: physchen@mail.ncku.edu.tw} }
\affiliation{Department of Physics, National Cheng-Kung University, Tainan 70101, Taiwan }

\author{ Takaaki Nomura \footnote{Email: nomura@kias.re.kr }}
\affiliation{School of Physics, Korea Institute for Advanced Study, Seoul 130-722, Republic of Korea}

\date{\today}

\begin{abstract}

ATLAS and CMS recently show the first results from run 2 of the Large Hadron Collider (LHC) at $\sqrt{s}=13$ TeV. A resonant bump at a mass of around 750 GeV in the diphoton invariant mass spectrum is indicated and the corresponding diphoton production cross section is around 3-10 fb. Motivated by the LHC diphoton excess, we propose that the possible resonance candidate is a Higgs singlet. To produce the Higgs singlet via gluon-gluon fusion process, we embed the Higgs singlet in the framework of vector-like triplet quark (VLTQ) model. As a result, the Higgs singlet decaying to diphoton final state is via VLTQ loops. Using the enhanced number of new quarks and new Yukawa couplings of the VLTQs and Higgs singlet, we successfully  explain the diphoton production cross section. We find that the width of the Higgs singlet is below 1 GeV, its  production cross section can be of order of  1 pb at $\sqrt{s}=13$ TeV, and the  branching ratio for it decaying to diphoton is around $0.017$ and is insensitive to the masses of VLTQs and new Yukawa couplings. We find a strong correlation between the Higgs Yukawa couplings to $s$-$b$ and $c$-$t$; the resulted branching ratio for $t\to c h$ can be $1.1\times 10^{-4}$ when the constraint from $B_s$ oscillation is applied. With the constrained parameter values, the signal strength for the SM Higgs decaying to diphoton is $\mu_{\gamma\gamma}< 1.18$, which is consistent with the current measurements at ATLAS and CMS.

\end{abstract}

\maketitle

\section{Introduction}

A scalar resonance with a mass of around 125 GeV was first discovered in the diphoton invariant mass spectrum and four-lepton channel at  the ATLAS~\cite{:2012gk} and CMS ~\cite{:2012gu} experiments. The scalar particle is identified as the Higgs boson which is responsible for the electroweak symmetry breaking (EWSB) in the standard model (SM). 

With $\sqrt{s}=13$ TeV, ATLAS and CMS recently report the first results from  run 2 of the Large Hadron Collider (LHC) and both experiments show a moderate bump at around $750$ GeV in the diphoton invariant mass spectrum, where ATLAS (CMS) employs $3.2(2.6)$ fb$^{-1}$ of data \cite{ATLAS-CONF-2015-081, CMS:2015dxe}.  With the narrow width approximation, the ATLAS and CMS experiments show that the local significances for the diphoton excess are $3.6\sigma$ and $2.6\sigma$ while the global significances are $2.0 \sigma$ and $1.2 \sigma$, respectively. 

The earlier search for diphoton resonances  was performed by ATLAS~\cite{Aad:2014ioa} and CMS~\cite{Khachatryan:2015qba} at $\sqrt{s}=8$ TeV. Although ATLAS found nothing exotic events, CMS indicated some excess at the diphoton invariant mass of  around $750$ GeV. 
Combined with the earlier diphoton excess, the local (global) significance at the CMS becomes $3.0(1.7)\sigma$. 
Motivated by the intriguing diphoton excess, the resonance candidates are proposed~\cite{Harigaya:2015ezk,Mambrini:2015wyu,Backovic:2015fnp,Angelescu:2015uiz,Nakai:2015ptz,Knapen:2015dap,Buttazzo:2015txu,Pilaftsis:2015ycr,Franceschini:2015kwy,DiChiara:2015vdm,Higaki:2015jag,McDermott:2015sck,Ellis:2015oso,Low:2015qep,Bellazzini:2015nxw,Gupta:2015zzs,Petersson:2015mkr,Molinaro:2015cwg,Dutta:2015wqh,Cao:2015pto,Matsuzaki:2015che,Kobakhidze:2015ldh,Martinez:2015kmn,Cox:2015ckc,Becirevic:2015fmu,No:2015bsn,Demidov:2015zqn,Chao:2015ttq,Gopalakrishna:2015dkt,Curtin:2015jcv,Fichet:2015vvy,Bian:2015kjt,Chakrabortty:2015hff,Ahmed:2015uqt,Agrawal:2015dbf,Csaki:2015vek,Falkowski:2015swt,Aloni:2015mxa,Bai:2015nbs}.

 Taking the CMS combined results as an illustration, if a resonance  at $750$ GeV exists, the cross section for the process $pp\to X \to \gamma\gamma $ should be around 3-10 fb, where $X$ denotes the unknown resonance. If we assume that the resonance is a scalar and its couplings to the SM fermions and gauge bosons are similar to the SM Higgs,  one can see that  the production cross section  times the branching ratio (BR) for $X\to \gamma\gamma$ decay is $\sigma(pp\to X) \cdot BR(X\to \gamma\gamma) \approx 1.33 \times 10^{-4}$ fb at $\sqrt{s}=13$ TeV~\cite{Heinemeyer:2013tqa}. The diphoton production cross section is too small to explain the excess. Therefore, if the resonance is a scalar boson, the main problem is that  $BR(X\to \gamma\gamma)$ should be of order of $10^{-3}$ when $\sigma(pp\to X)$ is  around 1 pb. 

Since the heavy scalar or pseudoscalar boson in an ordinary two-Higgs-doublet model (2HDM) can couple to the weak gauge bosons and/or   the SM fermions at the tree level, it may be difficult to escape  the problem of small $BR(X\to \gamma\gamma)$. To avoid this small BR, we propose that the new resonance candidate is a Higgs singlet ($S$) in which it does not couple to the gauge bosons and  the SM fermions at the tree level.  In order to produce this singlet by gluon-gluon fusion (ggF) process, we introduce new exotic quarks with masses of 1 TeV to  couple to the Higgs singlet.  

The proposal is motivated from the  loop induced effective interaction of the SM Higgs coupling to gluons, expressed as \cite{Gunion:1989we, Carmi:2012yp, Carmi:2012in, Jaeckel:2012yz}:
 \be
 {\cal L}_{hgg} = \frac{\alpha_s}{12\pi} \frac{y_t}{\sqrt{2} m_t} n_F h G^a_{\mu \nu}  G^{a\mu\nu}\,, \label{eq:Lggh}
 \ed
where $y_t$ is the top-quark Yukawa coupling, $v$ is the vacuum expectation value (VEV) of the SM Higgs, the mass of top quark is determined by $m_t = y_t v /\sqrt{2}$ , $n_F$ is the number of possible heavy quarks in the loop, and $n_F=1$ in the SM.  From Eq.~(\ref{eq:Lggh}),   it can be seen that the $h$ production cross section by ggF process can  be enhanced by the  Yukawa coupling and   the number of heavy quarks. For illustration, if we pretend $n_F= y_t=3$, $m_t=1$ TeV and $m_h =750$ GeV, the $h$ production cross section can reach ${\cal O}(1)$ pb. That is, the production cross sector for the  proposed Higgs singlet can easily reach the level of pb  when the number of new exotic quarks and Yukawa coupling are taken properly. Since the Higgs singlet can only decay via the new quark loops in which $S\to gg$ is the dominant decay channel; therefore, the  BR for the $S$ decaying to diphoton  can be naively estimated by $\Gamma(S \to \gamma\gamma)/\Gamma(S\to gg) \sim  32/256\, Q^4_F\, N^2_c /n^2_F\, \alpha^2/\alpha^2_s  \approx 3.0\times 10^{-3}$, where $Q^2_F$ is the sum of  squared electric charges  of new quarks inside loop and we take $Q^2_F=2$ and  $n_F=3$ as the example. Clearly, the resulted $\sigma(pp\to S\to \gamma\gamma)$ can match with the measurements from the LHC run 2. 

In order to establish a model that obeys the SM gauge symmetry, is anomaly free, possesses a scalar with a mass of around 750 GeV, and naturally provides  larger $n_F$ and Yukawa couplings, we investigate the subject in the framework of vector-like quark (VLQ) model  with a heavy $SU(2)_L$ Higgs singlet.  The related studies with Higgs singlet and/or VLQs for explaining the diphoton excess can be referred to~\cite{Angelescu:2015uiz,Knapen:2015dap,Buttazzo:2015txu,Franceschini:2015kwy,McDermott:2015sck,Ellis:2015oso,Dutta:2015wqh,Kobakhidze:2015ldh,Martinez:2015kmn,Falkowski:2015swt}.  From a fundamental theoretical perspective, which is for resolving hierarchy issue, matter-antimatter asymmetry,  etc, the VLQs  are predicted by the  theories, such as Little Higgs models~\cite{ArkaniHamed:2002qy, Han:2003wu, Perelstein:2003wd, Schmaltz:2005ky}, composite Higgs models~\cite{Kaplan:1983sm,Agashe:2004rs,Contino:2006nn,Anastasiou:2009rv,Vignaroli:2012sf}, extra dimensions~\cite{Antoniadis:2001cv, Hosotani:2004wv}, and nonminimal supersymmetric SM~\cite{Moroi:1991mg,Moroi:1992zk,Babu:2008ge,Martin:2009bg,Graham:2009gy,Martin:2010dc}. 
 A Higgs singlet can  be also embedded  in these models.
For phenomenological study, we directly add the VLQs and a Higgs singlet to the  SM. 

Basically, there is no limit for the  possible representations of VLQs. If we consider the VLQs those which can only mix with the SM up or down type quarks, the possible representations are singlet, doublet, and triplet~\cite{delAguila:2000rc, Okada:2012gy,Cacciapaglia:2012dd, Aguilar-Saavedra:2013qpa,Gopalakrishna:2013hua,Karabacak:2014nca,deBlas:2014mba,Cacciapaglia:2015ixa,Chen:2015cfa}. To avoid introducing too many VLQ states, we adopt the vector-like triplet quarks (VLTQs) in which each  triplet has three new quarks. In the base of gauge eigenstates, the introduced Higgs singlet  only couples to VLTQs.  

Since the introduced VLQs have different isospins from the SM quarks, therefore, the Higgs- and $Z$-mediated flavor changing neutral currents (FCNCs) occur at the tree level. Due to the new Yukawa couplings to VLQs and the SM quarks, besides the SM Higgs coupling to top-quark is modified, the SM Higgs couplings to VLQs are also induced. The SM Higgs production cross section and its decay to diphoton thus are changed. It is interesting to see  the influence of the model on the Higgs measurements and its implications at the LHC.  

The paper is organized as follows. In Sec.~II, we briefly introduce the model and present the new scalar potential, new Yukawa couplings, and new gauge couplings to the introduced vector-like quarks.  We study the properties of Higgs singlet and discuss its production and various decays  in Sec. III. In the same section, we also investigate the implications of new effects on the SM Higgs decay to diphoton, top FCNCs, and collider signatures. We give the conclusion in Sec.~IV. 

\section{Model}

  We extend the SM by including one real Higgs singlet $S$ and two VLTQs, where the representations of VLTQs in $SU(3)_c\times SU(2)_L \times U(1)_Y$ gauge symmetry are chosen as $(3,3)_{2/3}$ and $(3,3)_{-1/3}$ \cite{Okada:2012gy}.  Since the current Higgs measurements give a strict bound on the mixing angle between $S$ and the SM Higgs~\cite{Chen:2015cfa}, in order to suppress the mixing effect at the tree level, we impose a $Z_2$ discrete symmetry to the $S$ field, that is, $S\to -S$ under the $Z_2$ transformation. The scalar potential, obeyed the SM gauge symmetry and $Z_2$ symmetry, is expressed as:
 \begin{align}
 V(H, S) &= \mu^2 H^\dagger H + \lambda_1 (H^\dagger H)^2 + m^2_S S^2   + 
  \lambda_2  S^4 + \lambda_3 S^2 (H^\dagger H)\,. \label{eq:VHS}
 \end{align}
The representation of SM Higgs doublet is taken as:
  \be
H= \left(\begin{array}{cc}
 G^+    \\
\frac{1}{\sqrt{2}} ( v+ h + iG^0)     
\end{array}
  \right)\,,  
  \ed
where $G^+$ and $G^0$ are Goldstone bosons, $h$ is the SM Higgs field and $v $ is the VEV of $H$. The scalar potential in Eq.~(\ref{eq:VHS}) can not develop a non-vanished VEV for $S$ field when $\lambda_{2,3}>0$. Thus,  like the SM, $v=\sqrt{-\mu^2/\lambda_1}$ and $m_h= \sqrt{2 \lambda_1} v \approx 125$ GeV.   Due to the $Z_2$ symmetry,  $h$ and $S$ do not mix at the tree level and $m_S$ is the mass of $S$. We note that the $Z_2$ can be softly broken by the mass terms of VLTQs.

 The gauge invariant Yukawa couplings of VLTQs to the SM quarks,  the SM Higgs doublet, and the new Higgs singlet are written as: 
\be
-{\cal L}^{Y}_{\rm VLTQ} &=&  \bar Q_L {\bf Y_1} F_{1R} \tilde{H}  + \bar Q_L {\bf Y_2} F_{2R} H+   y_{1}  Tr(\bar F_{1L}  F_{1R} ) S
+  y_2  Tr(\bar F_{2L} F_{2R} ) S \non \\
&+& M_{F_1} Tr(\bar F_{1L}  F_{1R} ) + M_{F_2} Tr(\bar F_{2L}  F_{2R})+ h.c.\,,  \label{eq:yukawa}
\ed
where $Q_L$ is the left-handed SM quark doublet and  regarded as mass eigenstate before the VLTQs are introduced, all flavor indices are hidden,  $\tilde H =i \tau_2 H^*$,  and $F_{1(2)}$ is the $2\times 2$ VLTQ with hypercharge $2/3(-1/3)$. To keep the dimension-4 operator terms,  we require that $F_{1L, 2L}$ carry the $Z_2$ charge;  the representations of  $F_{1,2}$ in $SU(2)_L$ are expressed by:
  \be
F_{1} = 
\left(
\begin{array}{cc}
 U_1/\sqrt{2} & X    \\
 D_1 &  -U_1/\sqrt{2}     
\end{array}
\right)\,, \  F_{2} = 
\left(
\begin{array}{cc}
 D_2/\sqrt{2} & U_2    \\
 Y &  -D_2/\sqrt{2}     
\end{array}
\right)\,.
\ed
The electric charges of $U_{1,2}$, $D_{1,2}$, $X$ and $Y$ are $2/3$, $-1/3$, $5/3$ and $-4/3$, respectively. Therefore, $U_{1,2} (D_{1,2})$ can mix with up (down) type SM quarks.  The masses of VLTQs do not originate from the electroweak symmetry breaking. Due to the gauge symmetry, the VLTQs in the same multiplet state are degenerate and denoted by $M_{F_{1(2)}}$.  Since the mass terms of VLTQs do not involve $S$ field and the associated operators are dimension-3, therefore, the discrete $Z_2$ is softly broken by $M_{F_{1,2}}$ terms.

Since the $S$ field is a $SU(2)_L$ singlet, it can not directly couple to weak gauge bosons; however, the effective couplings to these gauge bosons can be induced through the VLTQ loops. Thus, it is necessary to study  the weak interactions of VLTQs.  We write the covariant derivative of $SU(2)_L\times U(1)_Y$ as: 
  \be
 D_\mu = \partial_\mu + i \frac{g}{\sqrt{2}} \left( T^+ W^+_\mu +T^- W^{-}_\mu \right) + i\frac{g}{c_W} \left(   T_3 -  s^2_W Q \right) Z_\mu+ i e Q A_\mu \,,
 \ed
where $W^{\pm}_\mu$, $Z_\mu$ and $A_\mu$ stand for the gauge bosons in the SM, $g$ is the gauge coupling of $SU(2)_L$, $s_W(c_W)=\sin\theta_W (\cos\theta_W)$, $\theta_W$ is the Weinberg angle, $T^{\pm} =T_1 \pm i T_2$, and the charge operator $Q = T_3 + Y$ with that $Y$ is the hypercharge of particle.  The generators of $SU(2)_L$ in triplet representation are set to be: 
 \be
  T_1=\frac{1}{\sqrt{2}}\begin{pmatrix}
  0 & 1 & 0 \\
   1 & 0 & 1 \\
    0 & 1 & 0 \\
  \end{pmatrix}\,,
~ T_2= \frac{1}{\sqrt{2}} \begin{pmatrix}     0 & -i & 0 \\
   i & 0 & -i \\
    0 & i & 0 \\
     \end{pmatrix}\,,~  T_3= \begin{pmatrix}
   1  & 0 & 0 \\
   0  & 0 &  0 \\
   0  & 0 & -1 \\
     \end{pmatrix}\,. 
 \ed
Accordingly, the gauge interactions of new quarks are summarized as:
 \be
 {\cal L}_{VFF} &=& -g \left[\left( \bar X \gamma^\mu U_1 + \bar U_1 \gamma^\mu D_1 + \bar D_2 \gamma^\mu Y + \bar U_2 \gamma^\mu D_2 \right) W^+_\mu + h.c. \right] \non \\
 &-& \left[ \frac{g}{c_W}  \bar F_1 \left( T^3 -s^2_W Q_1 \right) F_1 Z_\mu + e \bar F_1 \gamma^\mu Q_1 F_1 A_\mu  + ( F_1 \to F_2, Q_1 \to Q_2 ) \right]\,, \label{eq:VFF}
 \ed
where the alternative expressions for the VLTQs  are given by $F^T_1 = (X, U_1, D_1)$ and $F^T_2 = (U_2, D_2, Y)$ and the associated charge operators are diag$Q_1= (5/3, 2/3, -1/3)$ and diag$Q_2 = (2/3, -1/3, -4/3)$.  Since the left-handed and right-handed VLTQs have the same couplings to the gauge bosons, therefore, the vector-like quarks in Eq.~(\ref{eq:VFF}) are not separated by  their chirality.  
 
 \section{ Phenomena  of the Higgs singlet, the SM Higgs, and  the VLQs}

\subsection{Decays and production of the Higgs singlet}
After introducing the model, we analyze the production and decays of $S$ at $13$ TeV LHC. Since the $S$ mainly couples to the VLTQs, its production is through one-loop ggF process.  Thus,  the effective coupling induced from the VLTQ loops for $Sgg$  is formulated by:
  \be
{\cal L}_{Sgg} = \frac{\alpha_s}{8\pi } \left( \sum_{i=1,2} \frac{n_{F_i} y_i }{2m_{F_i}} A_{1/2}(\tau_i)  \right) S G^{a\mu \nu}G^a_{\mu \nu} \,,\label{eq:LggS}
 \ed
where $n_{F_i}= 3$ is the number of  VLTQs in the triplet state $F_{i}$ and the loop function is
 \begin{equation}
A_{1/2}(\tau) =  2 \tau [1+(1-\tau) f(\tau)^2]
  \end{equation}
  with $\tau= 4 m_{F}^2/m_S^2$ and $f(x)=\sin^{-1}(1/\sqrt{x})$.  Accordingly, the partial decay width for $S\to gg$ is derived by:
  \be
\Gamma(S \rightarrow gg)  =\frac{ \alpha_s^2 m_S^3 }{32 \pi^3}  \left|\sum_{i=1,2}\frac{n_{F_i} y_i}{2 m_{F_i}}    A_{1/2}(\tau_i) \right|^2\,. 
\label{eq:GHgg}
\ed
With the electromagnetic interactions in Eq.~(\ref{eq:VFF}),  the partial decay width via the VLTQ loops for $S\to \gamma\gamma$ is obtained as~\cite{Chen:2015cfa}: 
  \be
  \Gamma(S\to \gamma\gamma) = \frac{\alpha^2 m^3_S}{256 \pi^3}  \left| \sum_i \frac{y_i Q^2_{F_i} N_c}{m_{F_i}} A_{1/2}(\tau_i) \right|^2\,,
  \ed  
where $N_c=3$ is the number of colors and $Q_{F_i}$ stands for the total electric charge of triplet $F_i$. The partial decay width for $S\to Z\gamma$  can be formulated as:
\be
\Gamma(S\to Z\gamma) &=& \frac{N^2_c m^3_S }{32\pi} \left| A_F\right|^2  \left(1 - \frac{m_Z^2}{m_S^2}\right)^3 \,, \\
A_F &=& \frac{\alpha }{2\pi  s_W c_W} \sum_{i, f_i} \frac{-4 y_i Q_{f_i}}{m_{F_i}}( T^3_{f_i} - s^2_W Q_{f_i}) [I_1 ( \tau_{f_i}, \lambda_{f_i}) -I_2(\tau_{f_i}, \lambda_{f_i}) ]\,, \non 
\ed
where $i=1,2$, $f_i$ is the possible VLQ in $F_i$, $\tau_{f_i}= 4 m^2_{F_i}/m^2_S$, $\lambda_{f_i}= 4 m^2_{F_i}/m^2_S$, the summation is for $i$ and $f_i$, and the loop integrals are given as~\cite{Gunion:1989we}:
 \be
 I_1(a,b)&=& \frac{ab}{2(a-b)} + \frac{a^2 b^2}{2(a-b)^2} [ f(a)^2 - f(b)^2 ] + \frac{a^2 b}{(a-b)^2} [g(a)-g(b)]\,,\non \\
 I_2(a,b)&=&- \frac{a b}{2(a-b)} [ f(a)^2 - f(b)^2 ] \,, \non \\
 g(t)&=& \sqrt{t -1} \sin^{-1}(1/\sqrt{t})\,. 
 \ed
In order to calculate  the loop-induced $S\to W^+ W^-/ZZ$ decays,  we ignore the small effect from $m^2_{W(Z)}/m^2_{S}$ and the decay widths are derived as:
 \begin{align}
 \Gamma(S\to W^+ W^-) & = \frac{\alpha^2 m^3_S }{256 \pi^3} \left|\sum_{i=1,2}\frac{2 y_i N_c}{m_{F_i} s^2_W } A_{1/2}(\tau_i) \right|^2 \,, \non \\
 \Gamma(S\to ZZ)&=\frac{\alpha^2 m^3_S }{256 \pi^3} \left|\sum_{i,f_i}\frac{ y_i N_c (T^3_{f_i} -s^2_W Q_{f_i})^2 }{m_{F_i} s^2_W c^2_W} A_{1/2}(\tau_i) \right|^2\,.
 \end{align}
Since $Z_2$ is broken by the mass terms of VLTQs, $S\to hh$ decay can be induced at the loop level. The partial decay width is obtained as:
 \begin{align}
 \Gamma(S\to hh) & = \frac{m_S}{16 \pi } \lambda^2_{Shh} \sqrt{1-\frac{4m^2_h}{m^2_S}}\,,  \\
 \lambda_{Shh} &=  \frac{2 N_c y}{(4\pi)^2} \sum_i\left[ \frac{3m_S (Y^2_{i2} + Y^2_{i3})}{4m_{F_i}} I(m^2_S/m^2_{F_i})\right] \,, \non \\
 I(z)= & \int^{1}_{0} dx_1 \int^{x_1}_{0} dx_2 \frac{2 x_2 (x_2-x_1) + x_1/2}{x_1 + z  x_2 (x_2 - x_1)}\,.\non
 \end{align} 
 
 From Eq.~(\ref{eq:yukawa}), it can be seen that $Y_{1i}$ and $Y_{2i}$ can lead to new flavor mixing effects; as a result,  $S$ can decay to the SM quarks through the tree and loop diagrams. Nevertheless, the induced vertex is suppressed by $m_{q}/m_{F_i}$, where $m_q$ is the mass of the SM quarks. Therefore, $S\to t \bar t$ is the dominant decay mode.  In addition, the loop induced process has another  suppression factor $1/(4 \pi)^2$; therefore,  the partial decay width for $S\to t\bar t$ is of ${\cal O}(10^{-5})$ and negligible. Hence, we show the tree-induced partial decay width for $S\to t \bar t$ as:
  \begin{align}
  \Gamma(S\to t \bar t)& = \frac{m_S N_c }{8 \pi} \lambda_{Stt}^2\sqrt{1-\frac{4 m^2_t}{m^2_S}}\,, \\
  \lambda_{Stt} &= \frac{y_1 m_t v^2 Y^2_{13}}{4 m^3_{F_1}} + \frac{y_2 m_t v^2 Y^2_{23}}{2m^3_{F_2}}\,.\non
  \end{align}

\begin{figure}[hptb] 
\begin{center}
\includegraphics[width=70mm]{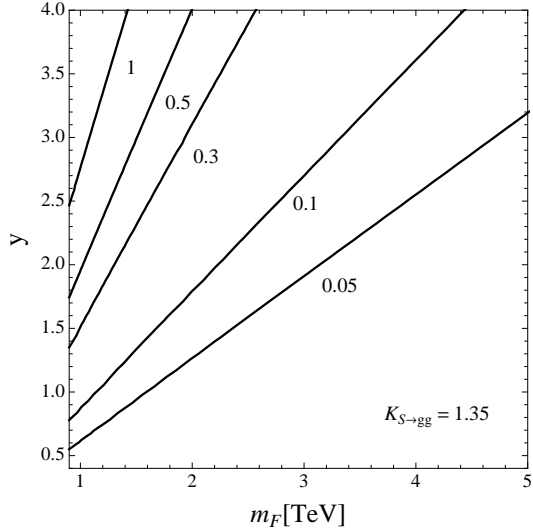} 
\caption{  Width of $S$ ( in units of GeV) as a function of $m_F$ and $y$.}
\label{fig:width}
\end{center}
\end{figure}

In order to perform the numerical analysis, without loss of generality we set $y_1 = y_2 =y$, $m_{F_1}=m_{F_2} = m_F$, and $m_S = 750$ GeV. To suppress $S$ decaying to the VLTQs, we require $m_F > m_S$. Thus, the main $S$ decay modes are $S\to f$ with $f=gg, W^+ W^-, ZZ, Z\gamma, \gamma\gamma, hh, t\bar t$ and the total width of the Higgs singlet is $\Gamma_S = \sum_f \Gamma(S\to f)$. In order to understand the total width of the Higgs singlet in the model, we present the contours for $\Gamma_S$ as a function of $m_F$ and $y$ in  Fig.~\ref{fig:width}, where the numbers on the plot denote the values of $\Gamma_S$ in units of GeV and the used K-factor is $K_{S\to gg}=1.35$~\cite{Djouadi:2005gi}. Clearly, without fine-tuning the Yukawa coupling $y$, the width of the Higgs singlet is below 1 GeV. Since ATLAS and CMS do not conclude the width of the resonance, a narrow or a wide width is possible. Therefore, the diphoton resonance in our model is a  narrow width scalar boson. Since loop-induced decays are all proportional to $y$, the BR for $S$ decay is independent of the Yukawa coupling. Numerically, we find that the BRs for $S\to gg/WW/ZZ/\gamma\gamma/Z\gamma/hh/t\bar t$ are also insensitive to the values of $m_F$ and they are: 
 \be
&& BR(S\to gg) \approx 0.673\,, ~ BR(S\to W^+ W^-) \approx 0.151\,, \non \\
  &&BR(S\to ZZ) \approx  0.095\,, ~ BR(S\to Z\gamma)\approx 0.039\,, \non \\
  && BR(S\to \gamma\gamma) \approx  0.0170\,, ~ BR(S\to hh) \approx 0.007\,, ~BR(S \to t\bar t) \approx 0.018\,,
  \ed
 where  $Y_{i2,i3}=1$ in $\lambda_{Shh}$ have been applied.  

\begin{figure}[hptb] 
\begin{center}
\includegraphics[width=70mm]{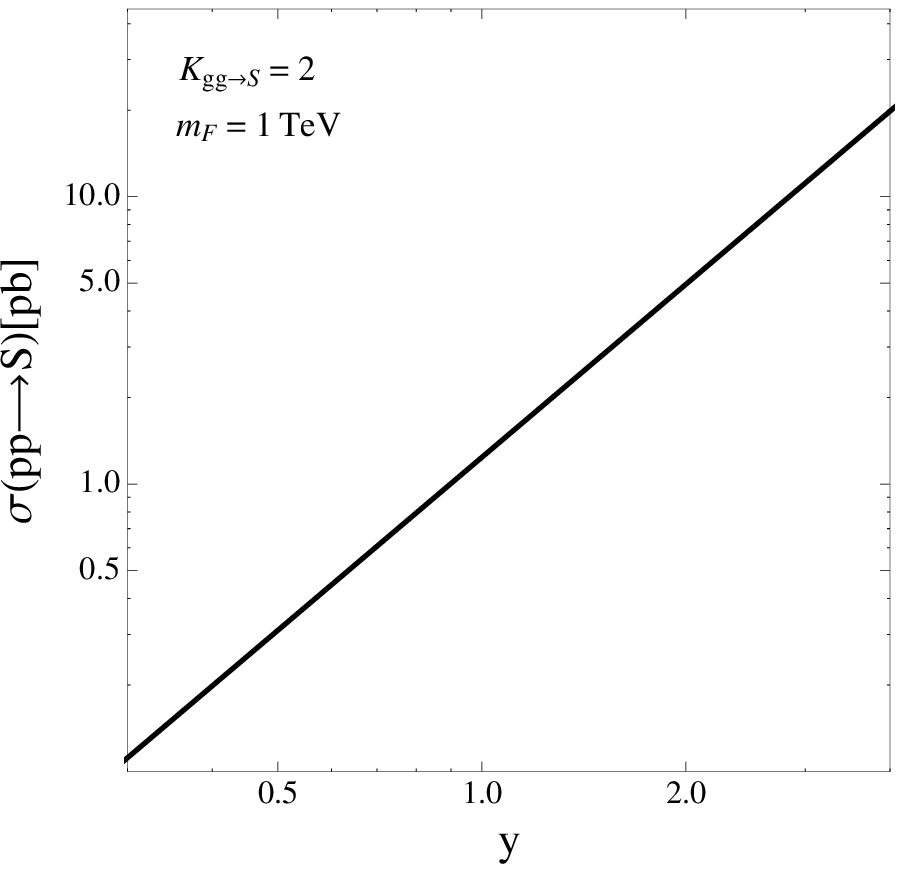}
 \includegraphics[width=70mm]{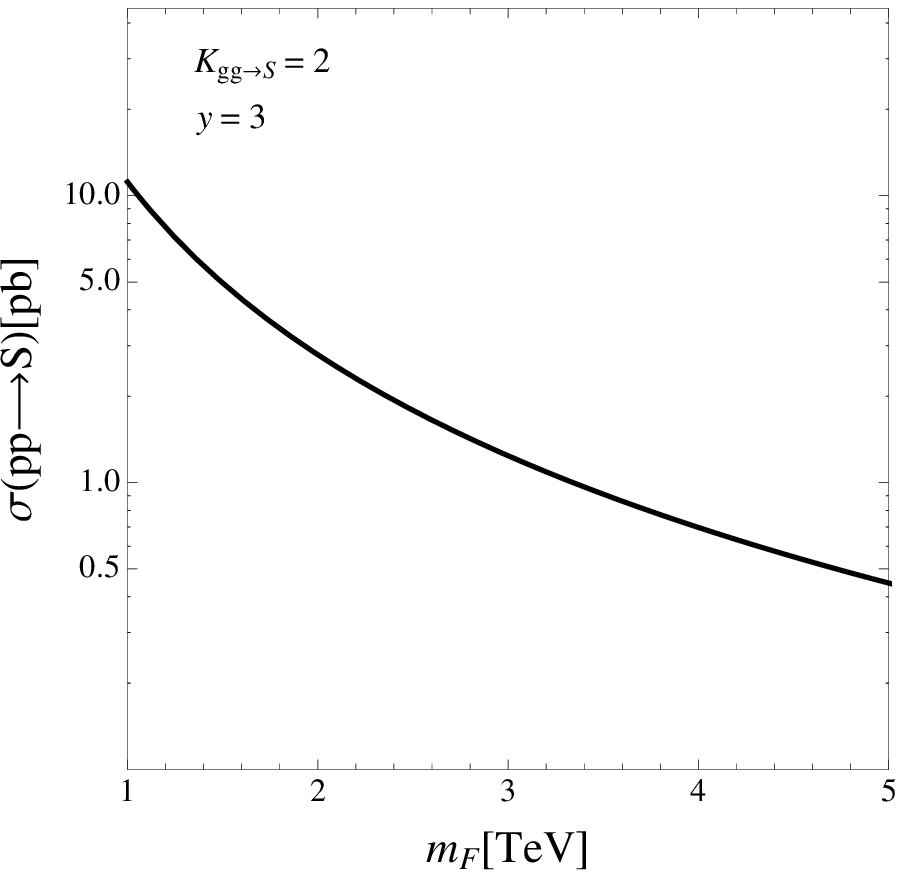}
\caption{  Higgs singlet production cross section (in units of pb) as a function of $y$ [left] and $m_F$ [right], where the K-factor is taken as  $K_{gg\to S}=2$. }
\label{fig:Xs}
\end{center}
\end{figure}

\begin{figure}[t] 
\begin{center}
\includegraphics[width=80mm]{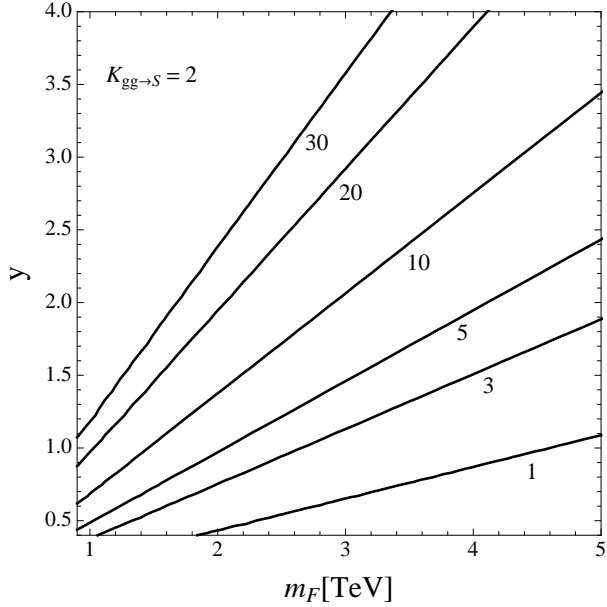}
\caption{  Contours for  $\sigma(pp\to S)\times BR(S\to \gamma \gamma)$ as a function of $m_F$ and $y$, where the numbers on the plot are in units of fb. }
\label{fig:Sgaga}
\end{center}
\end{figure}

In order to estimate the $S$ production cross section in $pp$ collisions  at $\sqrt{s}=13$ TeV, we implement the effective interaction of Eq.~(\ref{eq:LggS}) to the CalcHEP~\cite{Belyaev:2012qa}.  We display $\sigma(pp\to S)$ as a function of $y$ with $m_F=1$ TeV in the left panel of Fig.~\ref{fig:Xs}, where   {\tt CTEQ6L} PDF~\cite{Nadolsky:2008zw} is used and the K-factor is taken as $K_{gg\to S}=2$~\cite{Djouadi:2005gi}; in the right panel, we plot $\sigma(pp\to S)$ as a function of $m_F$ with $y=3$. It can be seen that the $S$ production cross section in pb can be achieved with that $m_F$ is at TeV scale. 
Combined the results of $\sigma(pp\to S)$ with those of $BR(S\to \gamma \gamma)$, we show the contours for $S$ production cross section times the BR for $S\to \gamma\gamma$ as a function of $m_F$ and $y$ in Fig.~\ref{fig:Sgaga}. From the plot, it is clear that with $y\sim O(1)$ and $m_F\sim O(1)$ TeV, the diphoton production cross section can match with the LHC diphoton excess. 
%

It is interesting to see if the bounds from other experimental upper limits can be satisfied when the values of parameters are fixed by the data of diphoton  excess. For performing such calculations, we require $\sigma(gg \to S \to \gamma \gamma) \approx  6$ fb at $\sqrt{s}=13$ TeV, which has combined the results of ATLAS and CMS~\cite{Franceschini:2015kwy,Ellis:2015oso}. Since the experimental upper bounds are measured  at $\sqrt{s}=8$ TeV, in order to get the theoretical results at $\sqrt{s}=8$ TeV, we simply divide the results at $\sqrt{s}=13$ TeV by the parton luminosity ratio $\sigma(gg \to S)_{\rm 13 TeV}/\sigma(gg \to S)_{\rm 8TeV} \approx 5$~\cite{Franceschini:2015kwy}. We present the results at $\sqrt{s}=8$ and $13$ TeV in Table~\ref{tab:1}.
It can be seen that our results are well below the experimental bounds.  It is worth mentioning that besides the $\gamma\gamma$ mode, from the table we see that the predicted $Z\gamma$ mode is very close to the current experimental bound. It can be a good candidate to test our model.

\begin{center} 
\begin{table}[t]
\begin{tabular}{|c|c|c|c|}\hline\hline  
Final sates & Constraints(8 TeV) & Our model (8 TeV) & Our model (13 TeV) \\ \hline 
$\gamma \gamma$ & $< 1.5$ fb~\cite{CMS:2014onr, Aad:2015mna} & $1.2$ fb & $6$ fb \\
$W^+ W^-$ & $< 40$ fb~\cite{Khachatryan:2015cwa,Aad:2015agg} & $11$ fb & $53$ fb \\
$ZZ$ & $< 12$ fb~\cite{Aad:2015kna} & $6.7$ fb & $34$ fb \\
$Z \gamma$ & $< 4.0$ fb~\cite{Aad:2014fha} & $2.8$ fb & $14$ fb \\
$jj$ & $\lesssim 2.5$ pb~\cite{Aad:2014aqa} & $47$ fb & $237$ fb \\
$hh$ & $< 39$ fb~\cite{ATLAS:2014rxa} & $0.49$ fb & $2.5$ fb  \\
$t \bar t$ & $< 550$ fb~\cite{Chatrchyan:2013lca} & $1.3$ fb & $6.4$ fb \\ \hline \hline
\end{tabular}
\caption{Comparisons of our results at $\sqrt{s}=8$ and $13$ TeV with the experimental bounds at $\sqrt{s}=8$ TeV, where the parton luminosity ratio $\sigma(gg \to S)_{\rm 13 TeV}/\sigma(gg \to S)_{\rm 8TeV} \approx 5$~\cite{Franceschini:2015kwy} has been applied and $\sigma(pp\to S\to \gamma\gamma)\approx 6$ fb is determined by the combination of ATLAS and CMS data at $\sqrt{s}=13$ TeV~\cite{Franceschini:2015kwy,Ellis:2015oso}.
}
\label{tab:1}
\end{table}
\end{center}

\subsection{ Top FCNCs, $h\to \gamma\gamma$,  and collider signatures}

We discuss other interesting processes  in the model below.
 From Eq.~(\ref{eq:yukawa}), it can be seen that after EWSB,  the FCNC interactions are induced  as:
  \be
  {\cal L}_{hQq} &=& \frac{Y_{1i}}{\sqrt{2}} (v+h)\left( \frac{1}{\sqrt{2} } \bar u_{Li} U_{1R} + \bar d_{Li} D_{1R} \right) \non \\
  &+& \frac{Y_{2i}}{\sqrt{2}} (v+h) \left( \bar u_{Li} U_{2R} - \frac{1}{\sqrt{2}} \bar d_{Li} D_{2R} \right)\,. \label{eq:hQq}
  \ed
Since the $D-\bar D$, $K-\bar K$, and $B_d-\bar B_d$ mixings make strict constraints on $Y_{i1}$,   the related effects  are small.  In the numerical analysis, we ignore their contributions. We find that  the FCNC interaction $hbs$ has a strong correlation to that $htc$ and can be written as:
 \be
 {\cal L}&=&- C_{sb} \bar s P_R b h - \frac{m_t}{m_b} C_{sb} \bar c P_R t h+ H.c.\,,  \label{eq:FCNC}\\
 C_{sb} &=& \frac{m_b}{4v} \left(2\zeta_{12} \zeta_{13} + \zeta_{22} \zeta_{23} \right)\non 
 \ed
with $\zeta_{ij}= Y_{ij} v/m_{F_i}$. With $\Delta m_{B_s}=1.1688\times 10^{-11}$ GeV, $f_{Bs}=0.224$ GeV, $m_{B_s (b)}=5.367 (4.6)$ GeV \cite{PDG},  the parameter $C_{sb}$ can be bounded as $|C_{sb}| < 5.2 \times 10^{-4}$.  If we take $\zeta_{12}\sim \zeta_{13} \sim \zeta_{22} \sim \zeta_{23} =\zeta$,  roughly  it can be seen $\zeta^2 < 0.036$. Using the coupling of $htc$ in Eq.~(\ref{eq:FCNC}), the decay rate for $t\to c h$ process is given by:
 \be
 \Gamma(t\to  ch) = \frac{m_t}{32\pi} \left|\frac{m_t}{m_b} C_{bs}\right|^2 \left(1 - \frac{m^2_h}{m^2_t} \right)^2\,.
 \ed
 With the width of top quark $\Gamma_t=1.41$ GeV~\cite{PDG} and the bound of $C_{sb}$, we get:
  \be
  BR(t\to c h) < 1.1\times 10^{-4}\,,
  \ed
 where the current upper limits from ATLAS and CMS are $0.46\%$~\cite{Aad:2015pja} and $0.56\%$~\cite{CMStch}, respectively. With a luminosity of ab$^{-1}$ at 14 TeV,  ATLAS estimates that the expected upper limit at the $95\%$ confidence level on the BR for $t\to c h$ decay can reach $1.5 \times 10^{-4}$ ~\cite{ATLAStch}. 
 
Due to the flavor mixing effects $Y_{1j}$ and $Y_{2j}$, the VLTQ loops can also contribute to the SM Higgs production cross section $\sigma(gg\to h)$ and the SM Higgs decay $BR(h\to gg/\gamma \gamma/Z\gamma)$. For illustrating the influence of VLTQs, we present the ratio of our model to the SM prediction for $pp \to h \to \gamma\gamma$ to be:
  \begin{align}
  \mu_{\gamma\gamma} &=  \frac{\sigma(pp\to h)_{\rm VLTQ}}{\sigma(pp\to h)_{\rm SM}} \frac{BR(h\to \gamma\gamma)_{\rm VLTQ}}{BR(h\to \gamma\gamma)_{\rm SM}} \non \\
  &\approx \left| 1 + \frac{3}{4} \zeta_{gg} \right|^2 \left| 1 + \frac{N_c A_{1/2}(x_F)\zeta_{\gamma\gamma}}{A_1(x_W) + 4/3A_{1/2}(x_t)} \right|^2 \,,
  \end{align}
where  we have adopted the limit $m_t, m_F \gg m_h$, $x_W=4 m^2_W/m^2_h$, $x_t=4m^2_t/m^2_h$, $x_F=4m^2_F/m^2_h$, $A_1(x_W) \approx -8.3$, $A_{1/2}(x_t) \approx 1.38$,
 \be
 \zeta_{gg}&=& \zeta^2_{12} + \zeta^{2}_{13}+\zeta^{2}_{22} +\zeta^2_{23}\,, \non \\
 \zeta_{\gamma\gamma}&=& \frac{Q^2_u + 2Q^2_d}{4} (\zeta^2_{12} + \zeta^2_{13}) +\frac{2Q^2_u + Q^2_d}{4} (\zeta^2_{22} + \zeta^2_{23}) \,,
 \ed
$Q_u=2/3$, and $Q_d=-1/3$. Using $\zeta^2_{ij} \sim \zeta^2 < 0.036$, we get
 $\mu_{\gamma\gamma} < 1.18$, where the contribution from VLTQs to $h\to \gamma\gamma$ is only $4\%$. The result is consistent with ATLAS of $\mu_{\gamma\gamma}=1.17\pm 0.27$~\cite{Aad:2015gba}   and CMS of $\mu_{\gamma\gamma}=1.13\pm 0.24$~\cite{CMS}.

We make some remarks on the constraints from the  electroweak precision measurements, such as the SM CKM matrix, $R_b$, and $R_c$. With the new Yukawa couplings in Eq.~(\ref{eq:yukawa}), the $3\times 3$ SM CKM matrix will be modified to be~\cite{Chen:2015cfa}:
 \be
 (V^{\rm SM}_{\rm CKM})_{i j} \to (V_{\rm CKM})_{i j}  + \frac{1}{2\sqrt{2}} \left( \zeta_{1i} \zeta_{1j} - \zeta_{2i} \zeta_{2j}\right)\,. \label{eq:MCKM}
 \ed
The modification will be smeared out if one takes $\zeta_{1i}\approx \zeta_{2i}$. Due to  the new flavor mixing effects, the $Z$ couplings to the SM quarks are also modified; therefore, the constraints from the electroweak precision measurements should be taken into account. Following Eqs.~(\ref{eq:VFF}) and (\ref{eq:hQq}), the new $Z$ couplings to the SM quarks are given by~\cite{Chen:2015cfa}: 
 \be
 {\cal L}_{Zq_iq_j} = -\frac{g}{8c_W} ( a_q \zeta_{1i} \zeta_{1j} - b_q  \zeta_{2i} \zeta_{2j}) \bar q_{iL} \gamma^\mu   q_{jL} Z_\mu \,, \label{eq:Zqiqj}
 \ed
where  $q_i$ denote the up- or down-type SM quarks, $a_u=b_d=1$,  $b_u=a_d=\sqrt{2}$, and only left-handed couplings are modified. With the scenario $\zeta_{i2}\approx \zeta_{i3}$, it can be seen that the changes of $Zc\bar c$ and $Zb\bar b$ couplings are the same in magnitude; here, we just examine the constraint from $R_b$. Using the results~\cite{Aguilar-Saavedra:2013qpa,Bamert:1996px}, we write 
 \begin{align}
 R_b &= R^{\rm SM}_b ( 1 - 3.56 \delta g^b_L)\,, \non \\
 \delta g^b_L &= \frac{1}{8}\left(\sqrt{2} \zeta^2_{13} - \zeta^2_{23} \right)\,.
 \end{align}
 With $R^{\rm exp}_b=0.21629 \pm 0.00066$~\cite{PDG} and  $R^{\rm SM}_b =0.21474$~\cite{Baak:2012kk},  the allowed range   in $2\sigma$ errors of data  for  $\zeta^2_{13}\approx \zeta^2_{23}=\zeta^2$ is $\zeta^2 \leq 0.066 $. The constraint is close to that from $B_s$ mixing.
 
 Finally, we  discuss the possible interesting signatures at the LHC. In the model, we introduce two top partners $U_{1,2}$, two bottom partners $D_{1,2}$, and two exotic quarks $X$ and $Y$ with the electric charges of $5/3$ and $-4/3$, respectively. The detailed studies of $U_{1,2}$ and $D_{1,2}$ can be referred to \cite{Aguilar-Saavedra:2013qpa,Gopalakrishna:2013hua,Karabacak:2014nca}; here we simply show what we find about the search for $X$ and $Y$. By using the CalcHEP,  the pair production cross sections for $X$ and $Y$ with $m_X=m_Y=1$ TeV at $\sqrt{s}=13$ TeV are $22$ fb; however, the single production cross sections of $X$ and $\bar Y$ by $W$-mediation can reach 100 fb when $\zeta=0.2$ is applied. With the scenario $\zeta_{12}\approx \zeta_{13}\approx \zeta_{22} \approx \zeta_{23}$, the main decay channels for $X$ and $Y$ in turn are $X\to W^+ (t, c)$ and $\bar Y\to W^+ (\bar s, \bar b)$  and each branching ratio is almost equal to 1/2. 
 Hence, the favorable channels to search for the single production of VLQs $X$ and $Y$ are
  \begin{align}
  pp & \to d X_{5/3} \to d W^+ c \,,\non \\
  pp &\to d X_{5/3} \to d W^+ t \to d W^+ (W^+ b)\,, \non \\
  pp & \to d Y_{4/3} \to d W^+ ( \bar s, \bar b)\,,
  \end{align}
where  the cross sections  can be  50 fb.  The detailed analysis and event simulations will be studied elsewhere.

 \section{Conclusion}
 
We employed a Higgs singlet $S$ to resolve the diphoton resonance with a mass of around 750 GeV, which is indicated by the ATLAS and CMS experiments when they analyzed the data from run 2 of the LHC at $\sqrt{s}=13$ TeV. In order to study the Higgs singlet production and decays, we embedded it to the VLQ model. Using the enhanced number of VLQs and new Yukawa couplings, we found that the $S$ production cross section can be of the order of 1 pb;  the BR for $S\to  \gamma\gamma$  is $0.017$ and is insensitive to new Yukawa couplings and masses of VLQs. As a result, $\sigma(pp\to S)\times BR(S\to \gamma\gamma)$ can match with the results of  3-10 fb, which are measured by the ALTAS and CMS experiments.
The width of the proposed Higgs singlet is below a few GeV; therefore our model is suitable for the analysis with  the narrow width approximation. 

We studied the implication on the FCNC process $t\to ch$ and found that  $BR(t\to ch)<1.1\times 10^{-4}$ when the data of $B_s$ oscillation  were included; with the same constrained value from $B_s$ mixing, we demonstrated that the signal strength for diphoton channel  is $\mu_{\gamma\gamma}<1.18$ and consistent with the current measurements in the ATLAS and CMS experiments. We examined the constraint from the precision measurement of $Z\to bb$ and the result is close to that from $B_s$ mixing. It is found that the single production cross sections for VLQs $X$ and $\bar Y$ can be over 100 fb and the dominant decay channels are $X\to W^+ (c ,t)$ and $\bar Y \to W^+ ( \bar s, \bar b)$.   \\
  
\noindent{\bf Acknowledgments} \\

 The work of RB was funded through the grant H2020-MSCA-RISE-2014 no. 645722 (NonMinimalHiggs) and 
was supported by the Moroccan Ministry of Higher Education and Scientific Research MESRSFC and  CNRST: "Projet dans les
domaines prioritaires de la recherche scientifique et du d\'eveloppement
technologique": PPR/2015/6.
The work of CHC was  supported by the Ministry of Science and Technology of  Taiwan, under grant  MOST-103-2112-M-006-004-MY3.  

\end{document}